\documentclass[preprint]{vgtc}               

\graphicspath{{figures/}{pictures/}{images/}{./}} 

\usepackage{times}                     
\usepackage{multirow}
\usepackage{mathptmx}                  

\onlineid{0}

\vgtccategory{Methods and Benchmarks}

\vgtcinsertpkg

\preprinttext{This manuscript was presented at VIS x GenAI, a workshop co-located with IEEE VIS 2025}

\newcommand{\gpt}{\texttt{GPT}\xspace}
\newcommand{\qwen}{\texttt{Qwen}\xspace}

\newcommand{\positive}{\texttt{positive}\xspace}
\newcommand{\negative}{\texttt{negative}\xspace}

\newcommand{\none}{\texttt{Novis}\xspace}
\newcommand{\nones}{\texttt{Novis$^*$}\xspace}
\newcommand{\amproof}{\texttt{AM}\xspace}
\newcommand{\circular}{\texttt{Circular}\xspace}
\newcommand{\force}{\texttt{Spring}\xspace}

\usepackage{xspace}
\newcommand{\rqone}{\textsf{RQ1}\xspace}
\newcommand{\rqtwo}{\textsf{RQ2}\xspace}

\usepackage{todonotes}
\usepackage{subcaption}
\usepackage{csquotes}
\usepackage[indentfirst=false]{quoting}
\quotingsetup{vskip=3pt}
\usepackage{booktabs}
\usepackage{enumitem}
\setlist[itemize]{itemsep=0pt, topsep=2pt}

\title{Visualization Biases MLLM's Decision Making in Network Data Tasks}

\author{Timo Brand\thanks{E-mail: firstname.lastname@tum.de} \textsuperscript{,}\thanks{These authors contributed equally.} %
\and Henry Förster\footnotemark[1] \textsuperscript{,}\footnotemark[2] %
\and Stephen G. Kobourov\footnotemark[1] %
\and Jacob Miller\footnotemark[1]}
\affiliation{\scriptsize Technical University of Munich, Heilbronn, Germany}

\teaser{
    \centering

        \begin{minipage}{0.15\linewidth}
            \includegraphics[width=\linewidth]{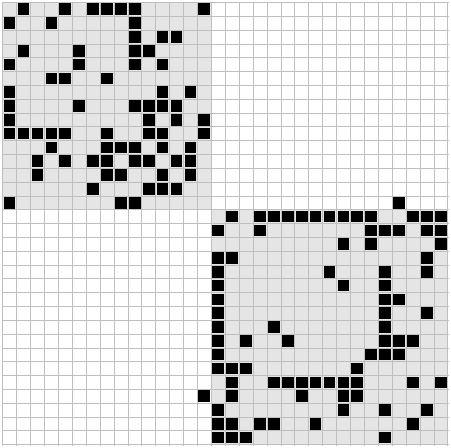} 
        \end{minipage}
        \ 
        \begin{minipage}{0.15\linewidth}
            \includegraphics[width=\linewidth]{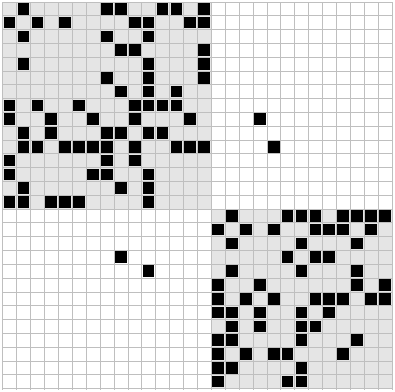} 
        \end{minipage}
        \hfill
        \begin{minipage}{0.15\linewidth}
        \centering
            \includegraphics[width=\linewidth]{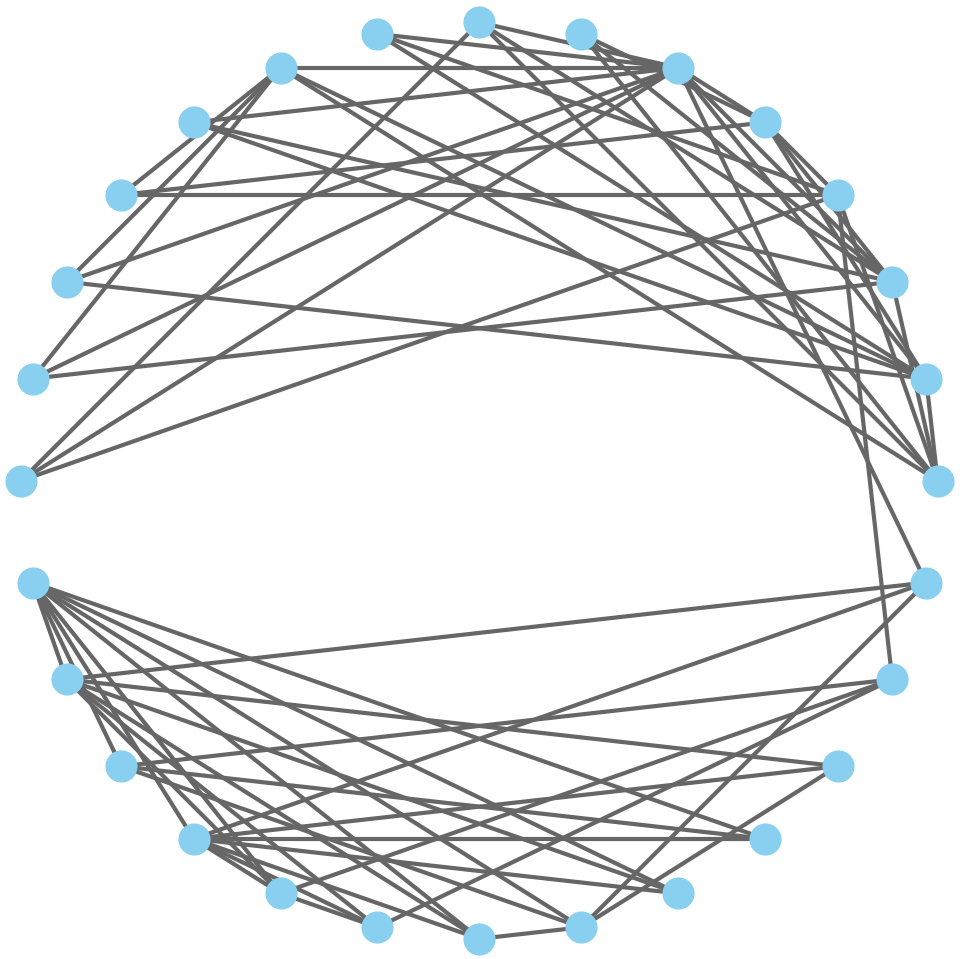}     
        \end{minipage}
        \ 
        \begin{minipage}{0.15\linewidth}
        \centering
            \includegraphics[width=\linewidth]{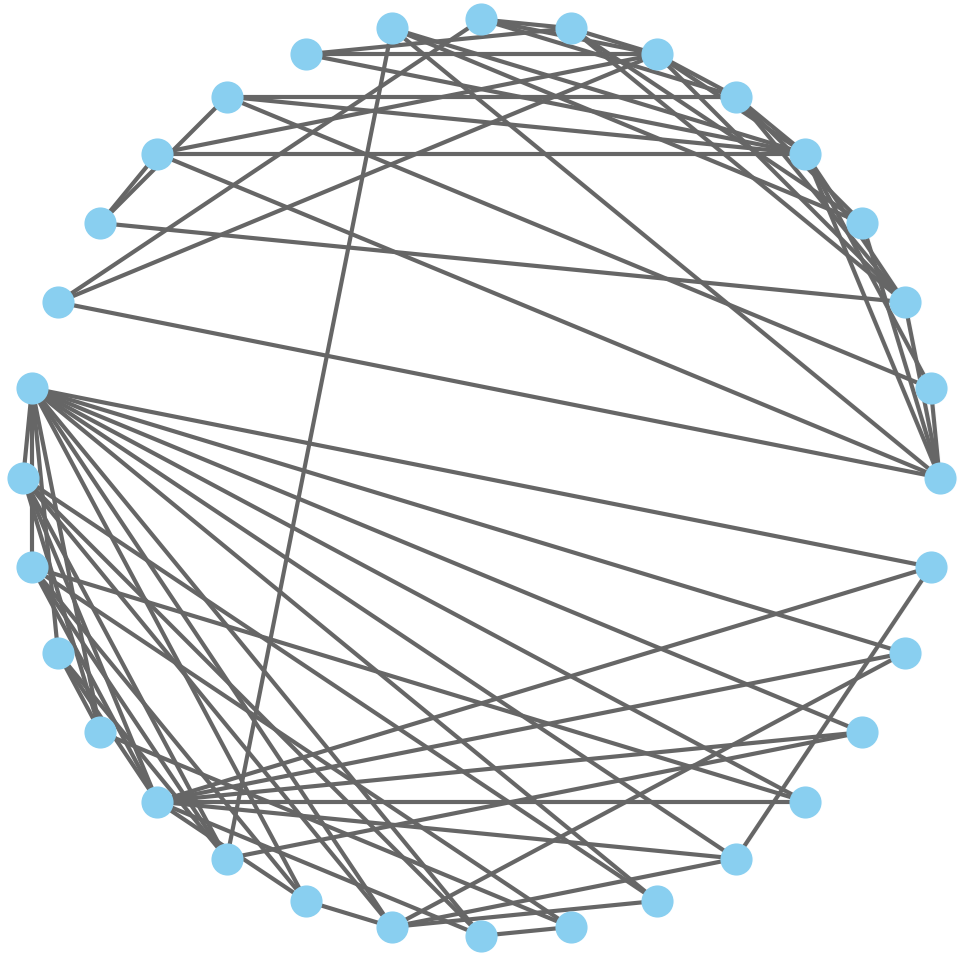}     
        \end{minipage}
        \hfill
        \begin{minipage}{0.105\linewidth}
            \includegraphics[width=\linewidth]{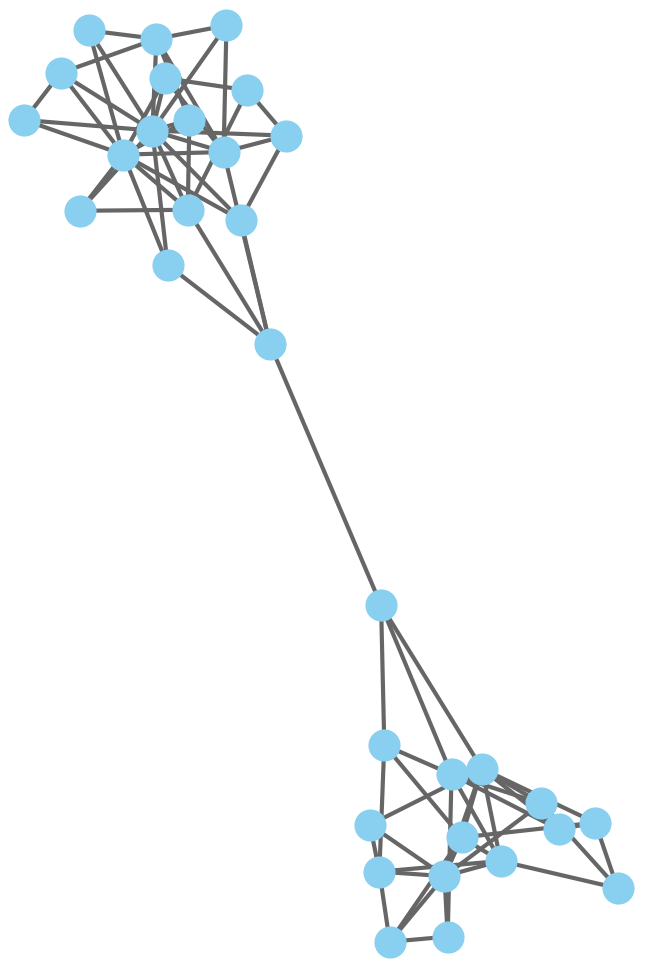}
        \end{minipage}
        \ 
        \begin{minipage}{0.195\linewidth}
            \includegraphics[width=\linewidth]{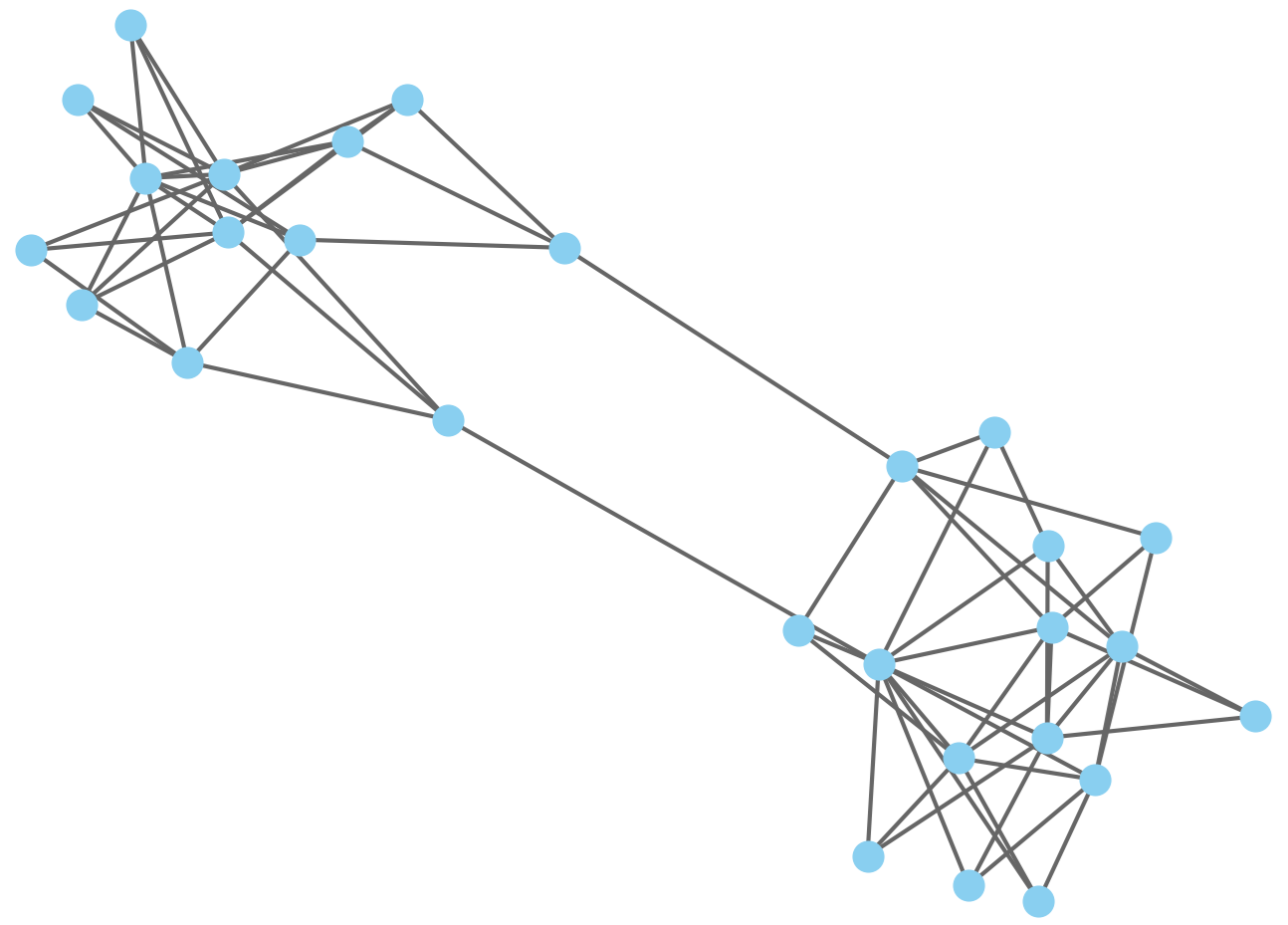}
        \end{minipage}

    \caption{Visualizations for two example graphs with a bridge and a 2-edge-connected example graph,
        using (from left to right) a pixel-based visualization of the adjacency matrix \amproof, a circular layout \circular, and a spring layout \force.
        Each block of two visualizations contains a \positive example of a graph with a bridge
        and a \negative example of a graph that is $2$-edge connected.
        Can you tell which one is which?}
    \label{fig:example-vis}

}

\abstract{
    We evaluate how visualizations can influence the judgment of MLLMs about the presence or absence of bridges in a network.
    We show that the inclusion of visualization improves confidence over a structured text-based input that could theoretically be helpful for answering the question.
    On the other hand, we observe that standard visualization techniques create a strong bias towards accepting or refuting the presence of a bridge --
    independently of whether or not a bridge actually exists in the network.
    While our results indicate that the inclusion of visualization techniques can effectively influence the MLLM's judgment without compromising its self-reported confidence,
    they also imply that practitioners must be careful of allowing users to include visualizations in generative AI applications so as to avoid undesired hallucinations.
} 

\keywords{network visualization, MLLM, bias, bridge,  visualization mirage,  visual proof}

\begin{document}


\firstsection{Introduction}

\maketitle

With recent developments in generative AI, \emph{large language models} (LLMs) are increasingly used as decision makers in practice.
Their nascent applications span a wide variety of domains, e.g., law~\cite{DBLP:journals/aiopen/LaiGWQY24},
finance~\cite{DBLP:conf/icaif/LiWDC23}, and healthcare~\cite{DBLP:journals/corr/abs-2311-05112}.
LLMs are now also beginning to be able to process multi-modal input.
In this context, it has been verified  that multi-modal large language models (MLLMs) possess some visualization literacy~\cite{DBLP:journals/tvcg/BendeckS25,10857634}.
Hence, MLLM decision makers might benefit from visualizations being provided in addition to the raw data,
akin to how human decision makers use visual analytics~\cite{DBLP:journals/jcst/SunWLL13} to support their decisions.
Another noteworthy aspect is that human users interacting with MLLM decision makers could attempt to influence
the MLLM's decision making process by augmenting data with visualizations, which may or may not be desired.

We see the need for assessing how MLLMs' decision-making processes can be guided by providing helpful visualizations.
In a  preliminary study, Förster et al.~\cite[supplemental material]{11027645} asked an MLLM whether a  network contained a Hamiltonian cycle.
The confidence of the MLLM's response could be improved when providing a \emph{visual certificate},
that is, a visualization highlighting the Hamiltonian cycle, compared to providing an adjacency matrix representation of the network as input.
This provides initial support for the hypothesis that visualization actually helps MLLMs in deriving correct solutions for tasks related to network data.
However, given their tendency to \enquote{hallucinate}, MLLMs might be prone to \emph{visualization mirages}~\cite{DBLP:conf/chi/McNuttKC20};
visualizations whose initial reading might support an erroneous hypothesis that is invalidated upon closer inspection.
In particular, the choice of visualization style might already create a bias.
We consider the following questions:
\begin{itemize}
    \item[\sffamily RQ1] \textit{Can the accuracy and confidence of MLLMs analyzing network data be improved when a suitable visualization is provided as part of the input?} 
    \item[\sffamily RQ2] \textit{Does the inclusion of visualization create a bias in the decision-making process of MLLMs and, if so, is such a bias dependent on the visualization style used?} 
\end{itemize}

We focus on answering \rqone and \rqtwo for a specific task -- determining if a network contains a \emph{bridge}; a single edge whose removal separates the network. 
If a network has no bridge, it is also called \emph{2-edge-connected}, as at least two edges must be removed before it becomes separated. 
In our experiments, we let a MLLM determine if a network contains a bridge and record the correctness and self-reported confidence.
We use standard visualization techniques, which we believe to be most likely  adopted by MLLM users.

Existing network visualization techniques are designed for humans, and it is unclear how much of their design principles apply to MLLMs.
Typically network visualizations focus on supporting overview tasks, displaying the entire data in an aesthetic and readable fashion~\cite{DBLP:journals/tvcg/AhmedLDKL22,DBLP:reference/crc/2013gd,DBLP:journals/tvcg/WangYHS24}.
Such visualizations display ground-truth structural properties of the underlying network faithfully~\cite{DBLP:conf/gd/NguyenEH12a}  facilitating free-form exploration by users.
They can be sufficient in supporting narratives in media by providing select views of the data~\cite{narratingnetworks}.
Adjacency matrices, circular layouts, and (force-directed) node-link diagrams all adhere to Munzner's expressiveness
and effectiveness principles~\cite{munzner2025visualization}, i.e.,
they show \textit{all and only} the data while making the representation \textit{effective} for the required task, e.g.,
identifying clusters, path-following, or identifying bridges; see \cref{fig:example-vis}.

\section{Experimental Setup}

In each trial, we present to one of the MLLMs, (\gpt\footnote{\small \url{https://platform.openai.com/docs/models/gpt-4o}}
or \qwen\footnote{\small \url{https://www.alibabacloud.com/help/en/model-studio/what-is-qwen-llm}}),
a small to medium-sized network described in text-form by an adjacency matrix and ask it whether the network contains a bridge. 
Each network is composed of two components $C_1$ and $C_2$ with either one or two links connecting them.
Hence, these edge(s) connecting $C_1$ and $C_2$ are exactly the links that may be the bridge in the network.
Our independent variable is the additional information given along with the network representation: 
We consider two text-based inputs, namely an unstructured adjacency matrix (i.e., a random permutation of the rows and columns)
and a structured adjacency matrix (i.e., a permutation of the rows and columns so that all nodes of $C_1$ precede all nodes of $C_2$).
In addition, we consider the setting where we provide a visualization in addition to the unstructured adjacency matrix,
which can be a pixel-based visualization of the structured matrix  or a standard node-link visualization (circular or force-directed layout).

\subsection{Generation of Stimuli}

\paragraph{Networks.} We generate a set of test-stimuli networks. 
For each network, we randomly sample  two subcomponents $C_1$ and $C_2$ using
the Barabasi-Albert model~\cite{doi:10.1126/science.286.5439.509}implementation in the \texttt{python} library \texttt{NetworkX}.
For the connectivity parameter, we use the value $3$ and we re-generate each component until we obtain a $3$-edge-connected one.
We create two sizes of components, small and large.
Small components have between 12 and 18 nodes, and large components have between 32 and 38 nodes, chosen uniformly at random.

We generate networks with a bridge by adding a single edge between a random node in $C_1$ and a random node in $C_2$.
For networks without a bridge, we do the same and choose another different node in each component and add an edge between them.
We say that the former ones are \positive and the latter ones are \negative instances. 
We generate 25 graphs for each of the 4 component size cases ($C_1$ small/large and $C_2$ small/large),
with a bridge and without one; i.e., in total 100 \positive and 100 \negative instances.

\begin{table*}[t]
\caption{Mean accuracy per model and configuration. 
The results are over all instances, only \positive, and only \negative instances.}
\label{tab:vis_acc_multicol}
\small
\centering
\begin{tabular}{l|ccc|ccc|ccc|ccc|ccc}
\toprule
& \multicolumn{3}{c|}{\none} 
& \multicolumn{3}{c|}{\nones} 
& \multicolumn{3}{c|}{\amproof} 
& \multicolumn{3}{c|}{\circular}
& \multicolumn{3}{c}{\force} \\
\multirow{-2}{*}{Model} & Total & Pos & Neg & Total & Pos & Neg & Total & Pos & Neg & Total & Pos & Neg & Total & Pos & Neg \\
\midrule
\gpt & 0.480 & 0.240 & 0.720 
     & 0.499 & 0.325 & 0.673 
     & 0.500 & 0.778 & 0.223 
     & 0.503 & 0.123 & 0.883 
     & 0.526 & 1.000 & 0.053 \\
\qwen & 0.502 & 0.825 & 0.178    
      & 0.499 & 0.589 & 0.408    
      & 0.490 & 0.940 & 0.036 
      & 0.469 & 0.019 & 0.937 
      & 0.509 & 1.000 & 0.000    \\
\bottomrule
\end{tabular}

\end{table*}

\begin{figure*}[t]
    \centering
    \includegraphics[width=0.49\linewidth]{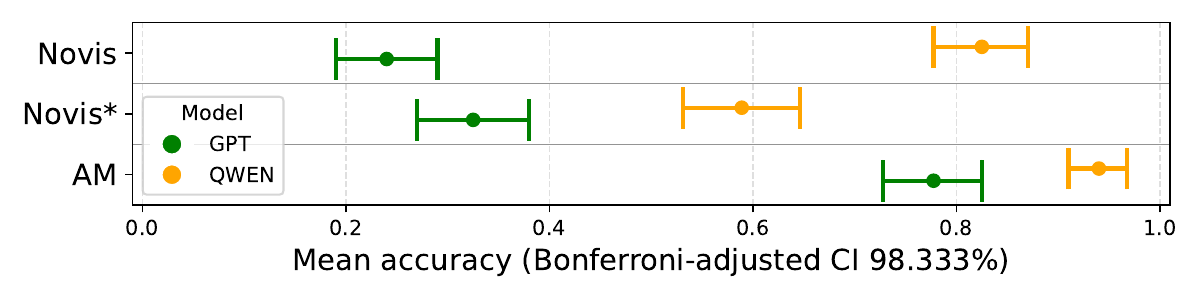}
    \includegraphics[width=0.49\linewidth]{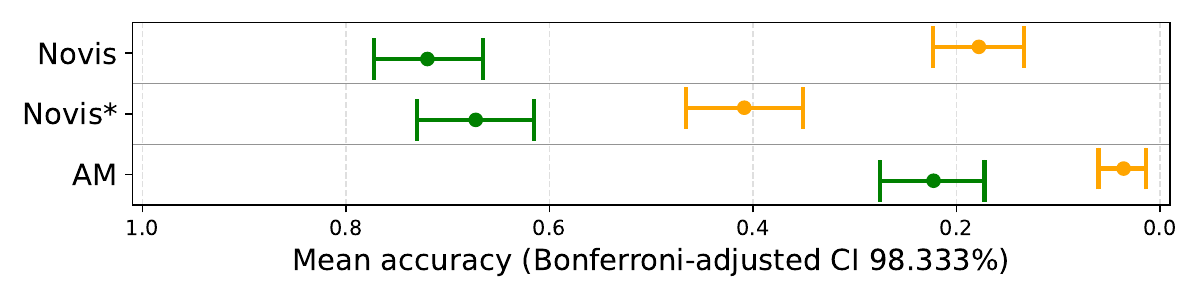}

    \caption{Confidence intervals for the results of the experiments using the adjacency matrix-based configurations \none, \nones and \amproof. 
    (left) shows the results for positive instances and (right) the negative instances.
    Note the inverted $x$-axis on the right.}
    \label{fig:new_ci}
\end{figure*}

\begin{figure*}[t]
    \centering
    \begin{minipage}[b]{0.65\linewidth}
    \includegraphics[width=0.49\linewidth]{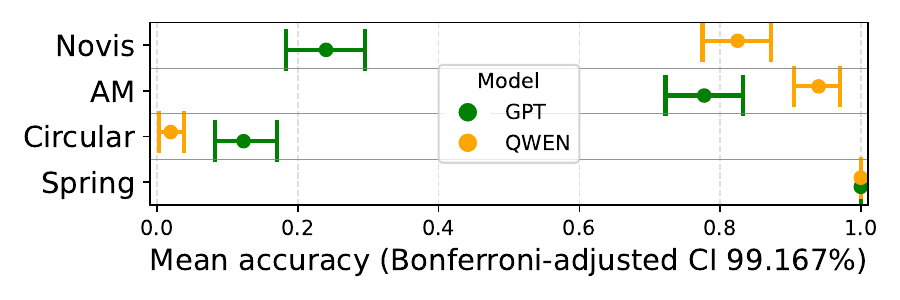}
    \includegraphics[width=0.49\linewidth]{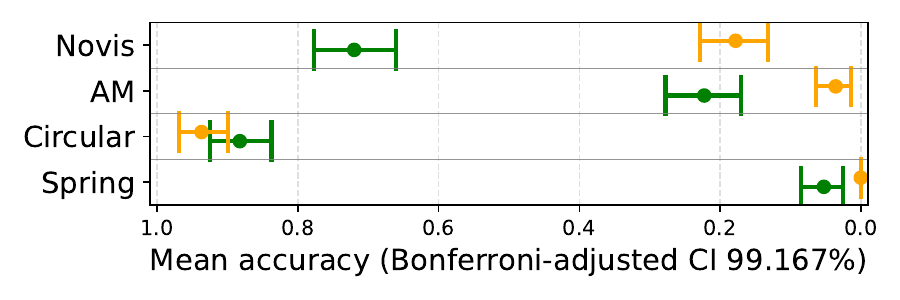}
    
    \includegraphics[width=0.49\linewidth]{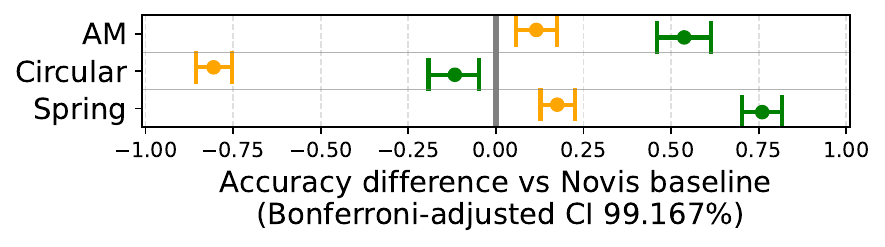}
    \includegraphics[width=0.49\linewidth]{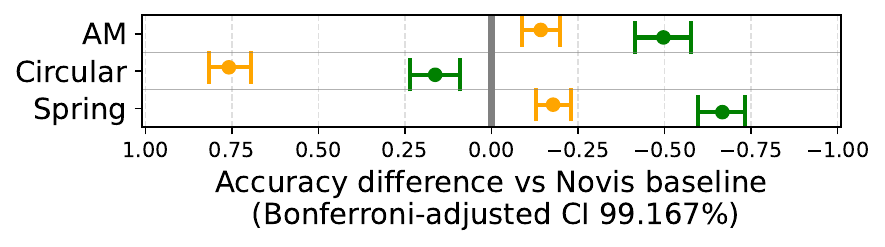}

    \caption{(top)
    Confidence intervals for the results of the experiments using the baseline \none and the visualization styles \amproof, \circular, and \force.
    (bottom)
    Confidence intervals for the difference of means test between the baseline at (0.0), and the visualization styles. 
    Intervals entirely to the right (left) of zero indicate significantly better (worse) performance than \none.
    Note the inverted $x$-axis on the right.}
    \label{fig:relative_baseline}
    \end{minipage}\hfill
    \begin{minipage}[b]{0.34\linewidth}
    \includegraphics[clip, trim=0cm 0.7cm 0cm 0cm, width=0.99\linewidth]{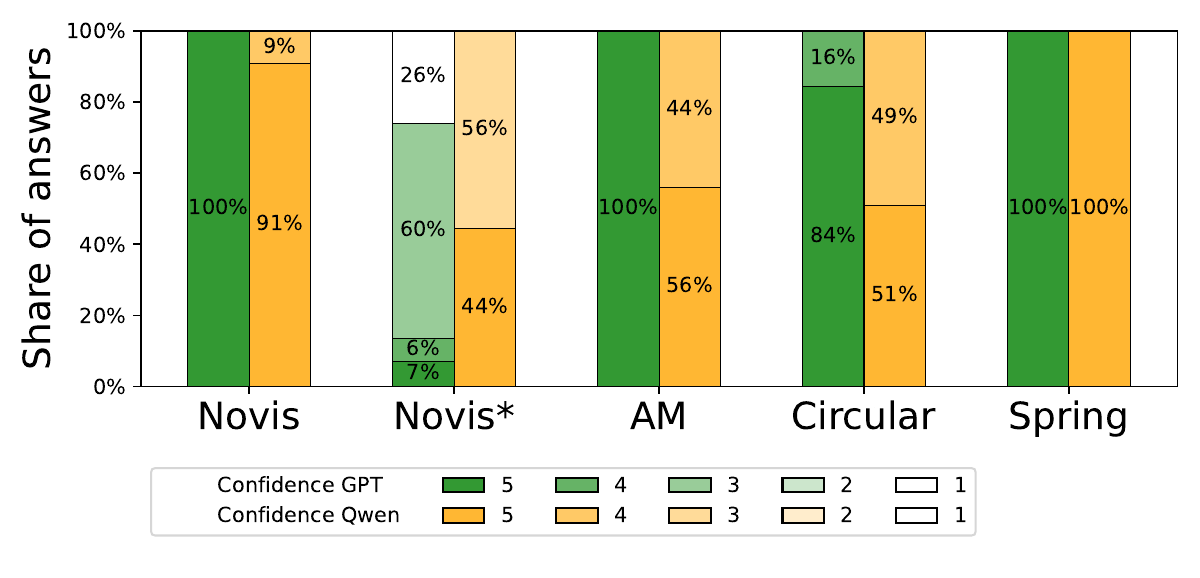}
    \includegraphics[width=0.99\linewidth]{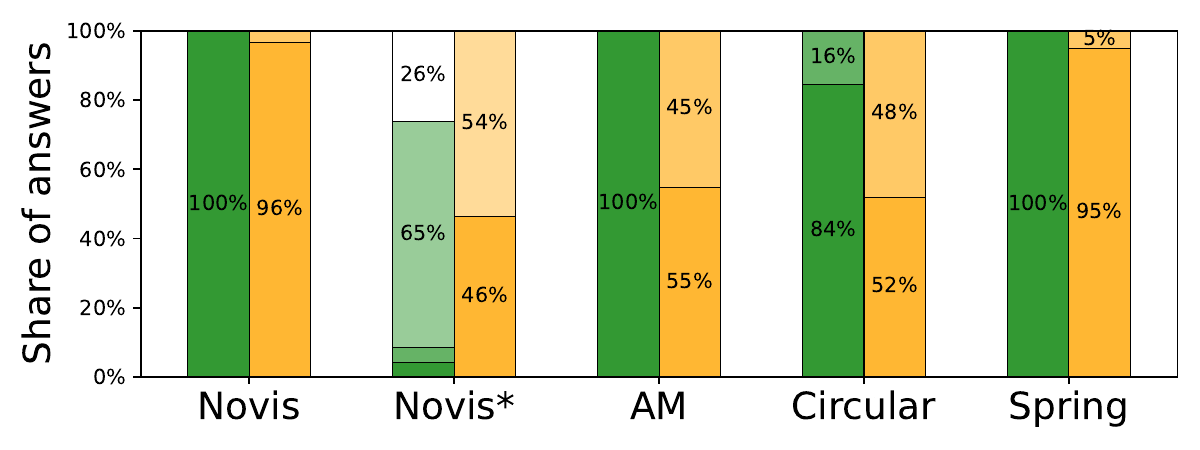}
    \caption{Share of answers with a certain confidence value 
     for (top) \positive and (bottom) \negative instances.}
    \label{fig:new_stacked_conf}
    \end{minipage}
\end{figure*}

\paragraph{Adjacency Information.}
We are interested in how providing a visualization of the graph impacts the performance of an MLLM in answering questions about graph properties.
As such, we supply a text representation of the graph and a possible layout.
As text representation, we chose adjacency matrices.
That is, we write for each vertex a separate row which contains a 1 if the vertex with the id of that column is adjacent to the vertex with the id of the row, and a 0 otherwise.
Before writing the adjacency matrix as text, we randomly permute the nodes to not leave any structure in the order of the rows and columns.
This data is later passed alongside a visualization to the MLLM.
Mainly to answer \rqone, we also experiment with two variants of text-only configuration:
\begin{itemize}
    \item \none: We provide only the text representation of the adjacency matrix.
    \item \nones: We provide a structured textual adjacency matrix representation, i.e., the nodes of $C_1$ come before the nodes of $C_2$ in the row/column ordering. 
    The order within each component is randomly chosen.
\end{itemize}

\paragraph{Visualizations}
In addition to the text-only configurations,
we also experiment with three configurations that pass a visualization of the graph alongside the text of the permuted adjacency matrix.

We generate the following layouts; see also \cref{fig:example-vis} (left to right):
\begin{itemize}
    \item \amproof: We provide a pixel-based visualization of the adjacency matrix where we sort the rows/columns so that all nodes of $C_1$ appear before all nodes of $C_2$, i.e., of the \nones configuration.
    Moreover, we enrich the visualization by shadings behind the square matrices representing $C_1$ and $C_2$. 
    \item \circular: We provide a circular layout generated with the function \texttt{circular\_layout} of \texttt{NetworkX}.
    We provide the network with a permutation of the nodes that separates the nodes of $C_1$ and $C_2$.
    As a result, the nodes of $C_1$ and $C_2$ are separated along the circle containing all the nodes, as in Fig.~\ref{fig:example-vis}.
    \item \force: We provide a force-directed layout generated using the \texttt{spring\_layout} function of \texttt{NetworkX}.
\end{itemize}

\subsection{Trial MLLM Prompts}

\paragraph{Prompt Structure.} 
We now describe how we performed a trial for the experiment. 
In each trial, we sent a prompt to the MLLM consisting of two parts, a \emph{system message} sent in the \texttt{system} role, and a \emph{trial instruction}  sent in the \texttt{user} role. All experiments have been performed as zero-shot experiments at temperature 0.0. That way, we aim to evaluate the baseline answers that an MLLM would produce when being exposed to the visual stimuli. Moreover, we did not allow MLLMs to access any APIs as depending on the practical use-case this behavior may be unwanted. In particular, this effectively prevents MLLMs from executing code.

\paragraph{System Message.} The system message explains the role of the MLLM using the following instructions:
\begin{quoting}
\small
\texttt{You are an expert graph-theory assistant.\\
    The user will provide two candidate statements (A and B) about a graph. Exactly one of them is true.\\
    Reply on ONE line in the form:\\
    ANSWER: $\langle$A/B$\rangle$ | CONFIDENCE: $\langle$1-5$\rangle$\\
    No extra text.}
\end{quoting}

The system message tells the MLLM that it is expected to perform well in the following trials, instead of attempting to emulate an average user. 
Second, it conveys that exactly one of the two given options is true and that it should report just one of them.

\paragraph{Trial Instruction.} The trial instruction consists of several parts. First, we formulate the question asked. There are two variants of the question (\texttt{\textbf{Q1}} and \texttt{\textbf{Q2}}), as we want to avoid a bias for the first or second option asked in the question:

\begin{quoting}
\small
\texttt{%
    {\textbf{Q1:}} Does the graph have an edge that, if removed, would disconnect it?\\
    {\textbf{Q2:}} Does the graph have no edge that, if removed, would disconnect it?}
\end{quoting}

We conclude the query with providing the possible answers:

\begin{quoting}
\small
\texttt{Choose exactly one option: A) [OPT\_A], B) [OPT\_B] Answer format: ANSWER: $\langle$A/B$\rangle$ | CONFIDENCE: $\langle$1-5$\rangle$}
\end{quoting}

\texttt{[OPT\_A]} and \texttt{[OPT\_B]} are placeholders for the following two answer options \texttt{\textbf{A1}} and \texttt{\textbf{A2}} whose order we make interchangeable:
\begin{quoting}
\small
\texttt{%
    \textbf{A1:} The graph does have such an edge.\\
    \textbf{A2:} The graph does not have such an edge.}
\end{quoting}

We create a trial for each combination of question and order of answers for each visualization-network pair.
After the query, we provide the adjacency matrix as text, following the word \enquote{\texttt{Adjacency:}}.
If the configuration includes a visualization, we append it as a PNG image generated with \texttt{NetworkX}
and saved using the function \texttt{savefig()} setting the parameter \texttt{dpi} to 300.
For each of the 200 stimuli and each combination of question and order of answers, we conduct a trial with each of the 5 configurations.

\section{Experimental Results}

We performed the experiments using the  MLLMs \gpt (\texttt{gpt-4.1-\allowbreak 2025-04-14}) and \qwen (\texttt{qwen2.5-vl-72b-instruct}).
For each trial, we recorded whether the MLLM's response was correct and the self-reported confidence in the range $1$ to $5$ provided as part of the MLLM's output. 
For each model, we compute the mean accuracy and Bonferroni-corrected confidence intervals over the set of configurations we are interested in. 
In the process, we apply the bootstrap statistical analysis method that takes a data
collection and creates many thousands of simulated samples
(of the same size as the original) by drawing from the original
data collection with replacement~\cite{tibshirani1993introduction}. 
For confidence scores, we record the proportion of responses for each reported confidence level.

More precisely, we make two comparisons: First, we compare the three adjacency matrix-based inputs \none, \nones and \amproof,
which differ in how structured and visual the information is provided.
Second, we investigate how the three visual configurations \amproof, \circular and \force perform in comparison to the base-line input \none.
Overall, the results are split over the \positive and \negative instances, 
to identify if any configuration causes a bias towards more often saying there is a bridge or there is none. 

\paragraph{Accuracy.} First observe that both models cannot effectively solve the task at hand,
independent of the input configuration, with an overall mean accuracy (including both \positive and \negative instances)
that is almost identical to a random coin flip; see \cref{tab:vis_acc_multicol}.
However, when evaluating \positive and \negative instances separately,
we see substantial differences in the performances of the different configurations -- %
all configurations perform well on either the \positive or the \negative instances, while they perform badly on the other set of instances.
We now evaluate this effect further.

Considering the  adjacent matrix-based methods \none, \nones and \amproof in \cref{fig:new_ci},
we observe that the MLLM's decision appears to be mainly dependent on the input configuration
instead of whether or not the instance actually is \positive or \negative.
This is evident from \cref{fig:new_ci}, where the accuracy is plotted increasing from left to right in the left subplot for \positive instances
and decreasing from right to left in the right subplot for \negative instances -- hence, in both figures,
a datapoint on the left-hand side (0.0 for the left and 1.0 for the right subplot) indicates that the MLLM assessed the input to be $2$-connected
whereas a datapoint on the right-hand side (1.0 for the left and 0.0 for the right subplot) indicates that the MLLM assessed the input to contain a bridge.
Since the means and confidence intervals in both subplots are almost identical,
we actually visually see that the MLLM's decision appears to be mainly based on the type of its input but not on the underlying data.

More precisely, with \amproof both models determine the network to contain a bridge in the majority of cases,
achieving high accuracy on the \positive and low accuracy on the \negative instances.
Curiously, for the text-based inputs \none and \nones, there is a discrepancy in the responses provided by \gpt and \qwen.
Namely, given only a text-based adjacency matrix, \qwen appears to be biased towards deciding that a given network contains a bridge whereas for \gpt the opposite is true.
Hence, for \none on the \positive instances, \qwen has significantly higher accuracy than \gpt, whereas on the \negative instances \gpt is more accurate than \qwen.
When structuring the data in \nones, we observe that both models become less stern in their decisions, with both models' accuracy getting closer to random guessing.
For \qwen this effect is more pronounced and statistically significant.

The difference between the accuracy for configuration \amproof and the accuracy for the text-based inputs \none and \nones is statistically significant for both models.
This is somewhat surprising as the configurations \nones and \amproof essentially communicate the same ordered data,
once as 0's and 1's in text form and once as either white or black pixels in the same matrix.
Despite that, their accuracy is significantly different for both \gpt and \qwen; for \qwen the effect size is even greater than in comparison to \none.

Next, consider how the visualizations \amproof, \circular and \force affect the accuracy compared to the baseline \none of passing only the adjacency matrix as text;
see \cref{fig:relative_baseline} (note that for negative instances, the accuracy axis again has increasing values from right to left).
Again, we observe a bias for both models to make the choice depending on the visualization style used.
In particular, \amproof and \force increase the probability for the MLLM to report that the network contains a bridge
whereas \circular increases the probability to receive an answer indicating a $2$-edge-connected network.
Hence, for positive instances, \force and \amproof achieve significantly higher accuracy than \none whereas \circular performs worse than \none.
In contrast, for \negative instances, with \circular significantly outperforming \none,
whereas \force and \amproof perform significantly worse than the text-based representation \none.

Based on \cref{fig:relative_baseline} (bottom), we observe that for \gpt, the bias created by \force is significantly greater than the one for \qwen.
In addition, \qwen achieves a higher divergence from \none based on \circular, whereas for \gpt the effect is stronger for \force and \amproof.
This difference is explained by the general bias of the corresponding \none evaluations; see again \cref{fig:relative_baseline} (top),
where we also see that \gpt and \qwen achieve similar response distributions for \force for \positive instances and for \circular for \negative instances.
In contrast, \amproof we observe different behaviors for \amproof.

\paragraph{Confidence.}

Both \gpt and \qwen  report the highest possible confidence score of $5$ in 78\% of the trials for \gpt  
and in 69\% of the trials for \qwen, independent of whether the instance is \positive or \negative. 
There is also little variance in their confidence, with regard to whether the models answered correctly or not: 
\gpt had a mean confidence of 4.5 (standard deviation of $1.06$) in both cases,
for \qwen it was 4.56 (standard deviation of $0.69$).

The share of answers with a certain confidence value for the different configurations is shown in \cref{fig:new_stacked_conf}
for \positive (top subplot) and \negative instances (bottom subplot).
Across both models, we observe no significant differences in reported confidence scores between \positive and \negative instances.

Regarding the adjacency matrix-based configurations, \none already leads to very high confidence for both models.
Surprisingly, passing the structured adjacency matrix text in \nones led to significantly lower confidence reported by both MLLMs.
If the structure however is encoded visually in \amproof, \qwen reports slightly higher confidence values compared to \none,
whereas for \gpt the behaviour is again quite different, always having confidence 5.

Finally, consider the two node-link styles.
The \circular layout seems to be the least convincing configuration across both models and introduces more doubt compared to \none,
with more answers given with confidence 4 (16\% for \gpt and 48--49\% for \qwen in \circular vs. 0\% for \gpt and 4--9\% for \qwen in \none).
The \force layout seems very convincing for both MLLMs, on the \positive instances both models always report a confidence of 5 in all trials,
on the \negative instances only \qwen reports a confidence of 4 on 5\% of the experiments.
For completeness, recall that \amproof performs slightly better than \circular,
achieving confidence score 5 for \gpt in 100\% of the trials whereas for \qwen the configuration's performance is more similar to \circular than to \none.

\section{Discussion}

We first evaluate \rqone:
\textit{Can the accuracy and confidence of MLLMs analyzing network data be improved when a suitable visualization is provided as part of the input?}

On all configurations, the MLLM's responses were rather influenced by a bias intrinsic to the selected MLLM
and to the chosen visualization style than by the factual data provided as part of the input.
In particular, this was evident from \cref{fig:new_ci,fig:relative_baseline} where we see that for all visualizations,
the distribution of choices by the MLLM was more or less the same, independent of whether or not we provided a \positive or a \negative instance as an input.
The visualizations \amproof and \force lead both models to respond that the network does contain a bridge,
even making them hallucinate the existence of a bridge for \negative instances very consistently for \qwen and still somewhat consistently for \gpt.
The opposite effect was observed for \circular where both models consistently reported that the network does not contain a bridge, even for \positive instances.
Again, the effect was slightly more pronounced for \qwen.
At this stage, we are tempted to refute that accuracy of the MLLM's responses can be improved
as the visualizations rather appear to steer the MLLM's judgment into some direction  in general.

Regarding the reported confidence values, surprisingly the structured text-based input \nones resulted in poorer confidence than the unstructured \none
even though \nones should be more useful for solving the task without executing code.
On the other hand,  encoding the same structure visually in \amproof did not come with any loss of confidence in \gpt and a less stark loss of confidence in \qwen.
Thus, it seems confidence can be improved if the data is to be structured as part of the input to facilitate the MLLM's judgment
(this may require more advanced prompting than in our experiment).
\medskip

Secondly, we evaluate \rqtwo: 
\textit{Does the inclusion of visualization create a bias in the decision-making process of MLLMs and,
    if so, is such a bias dependent on the visualization style used?}

We saw that each of the three visualization styles \amproof, \circular and \force creates a bias towards one or the other judgment,
which is consistent in its relation to the bias of \none over both models.
We also observed that a bias exists for the raw textual data \none as well which is slightly mitigated using structure in \nones.
In comparison, the bias by the visualizations is stronger than the bias for \none (see \cref{fig:relative_baseline})
and hence including a visualization as part of the input is far more likely to steer an MLLM's judgment -- independently of whether or not this achieves a desired effect.

\section{Conclusion}
In our experiments, both MLLMs appeared to be driven far more by the visualization design than by the underlying data; an effect that may seem to parallel human perception but with differences; e.g., the MLLM perceives no significant differences in the illustration pairs in \cref{fig:example-vis} while a human likely would. In fact, we intended creating purposefully misleading visualization mirages in a pilot study to make MLLMs draw wrong conclusions until we noticed that
two edges between two components in the \force configuration were already consistently and confidently interpret  as a single bridge.

For visualization research, our results indicate  that evaluating MLLMs as \enquote{human-like} readers can be problematic,
as although they might produce similar results in the aggregate, the fine-grained distinction tells a different story.
Understanding these differences is essential as MLLMs become more and more widely used tools for prototyping, studying,
and evaluating visualization techniques; a great avenue for future work.
For generative AI, the demonstrated increase in confidence scores when providing visualizations
in addition to raw data input may be a desirable effect in automated pipelines.
However, the fact that visualization itself appears to create strong biases indicates that strong caution has to be taken
when designing an application that steers MLLMs using visualization.

\paragraph{Limitations.} Our results have several limitations that must be taken into account to avoid over-generalizations.
We focused on a single task, the detection of a bridge.
Future experiments could investigate if similar effects can be observed for other network tasks.
Moreover, we restricted ourselves to two standard layout techniques with circular layout and a force-directed algorithm.
Other state-of-the-art algorithms might be taken into account in future studies.

We investigated both a recent commercial MLLM model and an open-source model and observed non-trivial differences between them.
Hence, it is not so clear if our results can be generalized to other MLLM models.
Finally, our experiments used zero-shot prompting and did not allow for tool access.
Potentially, a more fine-tuned prompt, using, for instance, few-shot prompting or RAG, may yield results that are not influenced by the MLLM's bias.

\bibliographystyle{abbrv-doi}
\bibliography{bibliography}
\end{document}